\documentclass[aps,prb,twocolumn,superscriptaddress,reprint,showpacs, bibliography]{revtex4-1}

\usepackage{graphicx}
\usepackage{amsmath}
\usepackage{amssymb}
\usepackage[usenames,dvipsnames]{color}
\usepackage{pdfcolmk}

\usepackage[utf8]{inputenc}
\usepackage{color,xcolor}

\usepackage{hyperref}
\hypersetup{colorlinks=true,citecolor=blue,linkcolor=blue,filecolor=blue,urlcolor=blue,pdfstartview=FitH,pdfpagemode=UseNone}

\usepackage{array}
\newcolumntype{L}[1]{>{\raggedright\let\newline\\\arraybackslash\hspace{0pt}}m{#1}}
\newcolumntype{C}[1]{>{\centering\let\newline\\\arraybackslash\hspace{0pt}}m{#1}}
\newcolumntype{R}[1]{>{\raggedleft\let\newline\\\arraybackslash\hspace{0pt}}m{#1}}

\bibpunct{[}{]}{,}{n}{}{}

\begin{document}

\title{Disordered Si:P nanostructures as switches and wires for nanodevices}

\author{Amintor Dusko}
\email{amintor.dusko@gmail.com}
\affiliation{Instituto de F\'isica, Universidade Federal Fluminense, 24210-346 Niter\'oi, RJ, Brazil}

\author{Belita Koiller}
\affiliation{Instituto de F\'isica, Universidade Federal do Rio de Janeiro, Caixa Postal 68528, 21941-972 Rio de Janeiro, Brazil}

\author{Caio Lewenkopf}
\affiliation{Instituto de F\'isica, Universidade Federal Fluminense, 24210-346 Niter\'oi, RJ, Brazil}
		\date{\today}
		
\begin{abstract}

Atomically precise placement of dopants in Si permits creating substitutional P nanowires by design.
High-resolution images show that these wires are few atoms wide with some positioning disorder with respect to
 the substitutional Si structure sites. Disorder is expected to lead to electronic localization in one-dimensional (1D) - like structures.
Experiments, however, report good transport properties in quasi-1D P nanoribbons.
We investigate theoretically their electronic properties using an effective single-particle approach based on a linear
combination of donor orbitals (LCDO), with a basis of six orbitals per donor site, thus keeping the ground state donor
orbitals' oscillatory behavior due to interference among the states at the Si conduction band minima.
Our model for the P positioning errors accounts for the presently achievable placement precision allowing to study the
localization crossover.
In addition, we show that a gate-like potential may control its conductance and localization length, suggesting the possible use of Si:P nanostructures as elements of quantum devices, such as nanoswitches and nanowires.
\end{abstract}
\maketitle

\section{Introduction}
\label{sec:intro}

The approaching breakdown of Moore's law has triggered a strong research effort to avoid compromising the
miniaturization spiral in electronics. One of the promising strategies to keep it evolving consists in transferring
current device functionalities to nanostructures prepared with atomic-scale control.
Given the ubiquity of silicon integrated circuits presently in use, atomic implantation of dopants in Si hosts constitutes
a very attractive road towards achieving such structures.
This requires effective control of donor positioning at pre-assigned sites, {\it i.e.}, fabricating devices at the atomic level
by design \cite{Weber2012, Weber2014, Shamim2016}.

Reports of successful placement of P arrays in Si suggest that this arrangement could, in  principle, play the role of
nanowires connecting different components of nanodevices, similar to a metallic wire in regular chips \cite{Schofield2003,
 Weber2012, Fuechsle2012, Zwanenburg2013, Weber2014, Salfi2016, Shamim2016}.

The adequacy of P nanochains and nanoribbons in Si to serve as channels for electronic transport in devices raises some
questions. In principle, a perfectly ordered array does provide the desired connections. However, in real samples the positioning
uncertainties, inherent to the current fabrication processing standards, may spoil the desired conductance features:
Due to the well known property that electronic states in disordered one-dimensional (1D) materials are localized, disordered nanowires can
become insulators, with negligible electronic transport. Since the nanowires of  interest here are finite,  the transmission of electrons
is possible, as long as the electronic localization length is comparable or larger than the system length itself \cite{Dusko2016}.

Here we investigate these questions theoretically, modeling P nanochains and nanoribbons by a tight-binding description with
6 orbitals per P substitutional site, corresponding to the combinations of the 6 minima in the Si conduction band,
symmetrized  according to the tetrahedral crystal field potential at the donor site, see Appendix.  The sixfold degenerate levels split
into states that have the symmetry of the different irreducible representations of the T$_\textrm{d}$ group \cite{Kohn1957}.
This leads to a singlet with A$_1$ symmetry, a triplet with T$_2$ symmetry, and a doublet with E symmetry.
Starting from an ideal target configuration for the P sites, the actual positions are individually chosen according to a Gaussian
distribution of lattice positions centered at each target site.

In this multi-orbital scenario, we systematically study how the choice of the device geometry, namely, the interdonor
distance and the wire dimensions (width and length) affect the system's electronic conductance and localization.
In addition, we show that such generated nanostructures can serve as nanoswitches controlled by an external gate potential.

This paper is organized as follows: In Sec.~\ref{sec:models_and_methods} we summarize the theoretical LCDO scheme,
the atomistic model considered here, and the Landauer-B\"uttiker approach for quantum coherent transport.
In Sec.~\ref{sec:transport_and_placement} we outline the localization length calculation scheme and compare the main
features of the different disorder intensities scenarios.
In Sec.~\ref{sec:loc_length} we investigate  the sensitivity  of the localization length parameter to an external gate potential
and in Sec.~\ref{sec:conductance} we analyze the corresponding effects on the nanostructure conductance.
Our conclusions and summary are presented in  Sec.~\ref{sec:conclusions}.

\section{Model and Methods}
\label{sec:models_and_methods}

The full set of electronic states that describe mesoscopic nanostructures formed by donors in a Si host correspond
to a Hilbert space whose size is typically larger than $10^6$ atomic orbitals.
As demonstrated in Refs.~\cite{Dusko2016, Dusko2018}, the Hilbert space can be effectively represented
by a reduced basis formed by a Linear Combination of Donor Orbital (LCDO).
In this hybrid method each donor orbital is accounted for by a multi-valley central cell effective mass approach,
that incorporates the Si host effects in the donor orbital itself.

\begin{figure*}[t!]
		\includegraphics[clip,width=0.9\textwidth]{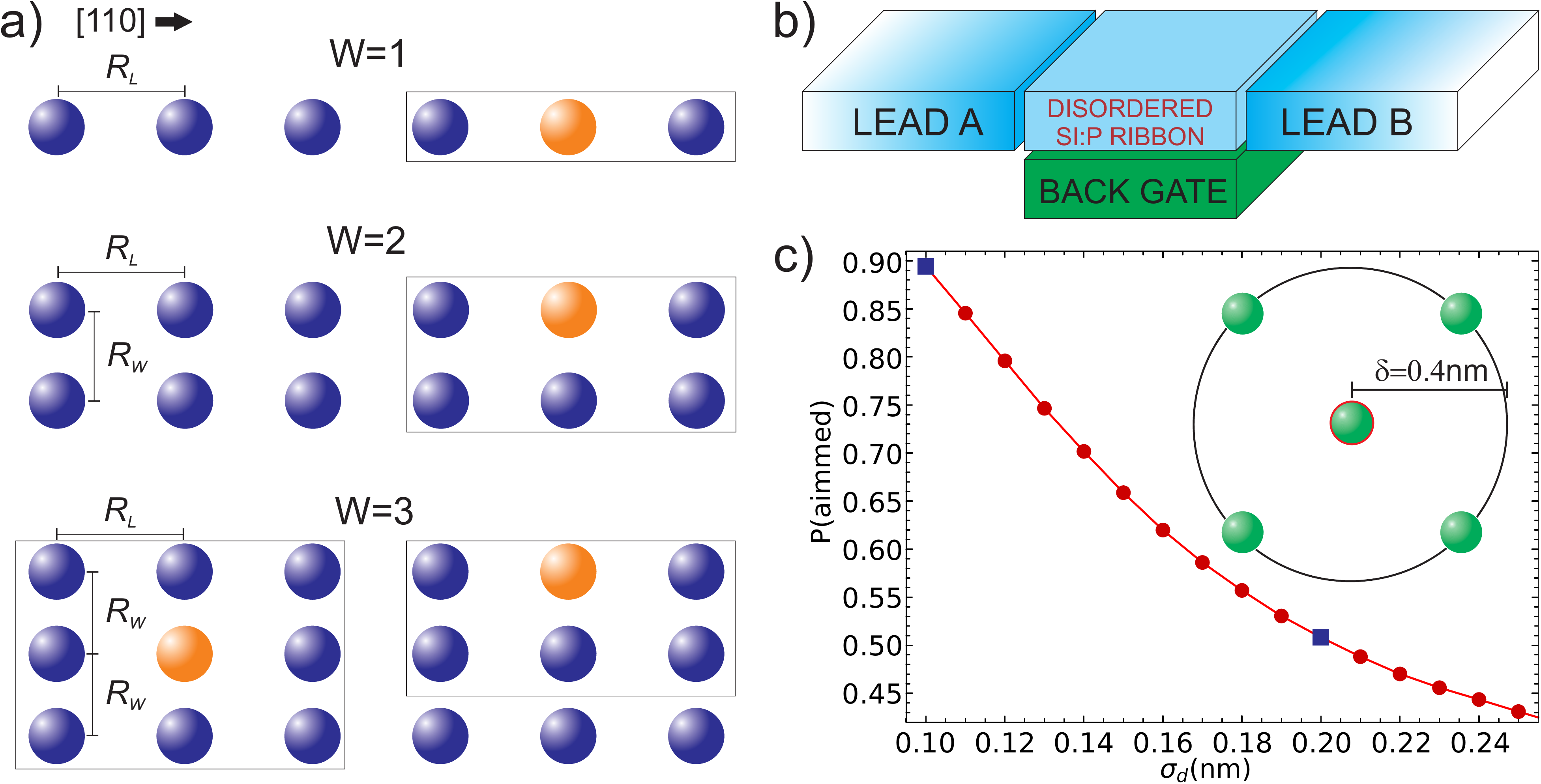}
		\caption{\label{fig:geometry}
(a) Target Si:P nanostructure fragment, along the Si [110] crystalline direction, for different widths $W$, specifying the
geometric parameters $R_W$ and $R_L$.
The rectangles define a neighboring region around a reference site (orange sphere).
The number of neighbors sites changes with $W$. For $W>2$ one can define edge and bulk sites, with
5 and 8 neighbors respectively.
(b) Sketch of the model system: disordered Si:P sample connected to semi-infinite leads with translational symmetry,
subjected to a back gate potential.
(c) Probability distribution $P$ of an implanted donor to occupy the aimed position as a function of $\sigma_d$.
Two cases are studied in this paper (blue squares):  $\sigma_d=0.1$ and $0.2$ nm, corresponding to a deposition
matching the aimed position 90\% and 50\% of the time, respectively.
Inset: Graphical representation of the disorder cutoff radius $\delta$.
Here green spheres indicate the Si structure and the red circled sphere is the target position.
}
\end{figure*}

We characterize the nanostructures by four geometric parameters, namely, width ($W$), length ($L$), transversal donor
distance ($R_W$) and longitudinal donor distance ($R_L$), see Fig.~\ref{fig:geometry}(a).
Considering the placement process to occur along the Si $\langle 110 \rangle$ direction, the target P donors form
a rectangular lattice with lattice parameters defined by $R_W$ and $R_L$.

The model multi-orbital Hamiltonian written in the LCDO basis~\cite{Dusko2016} reads
\begin{equation}\label{eq:hamiltonian}
  H  =\sum_{i,l}\varepsilon^{}_{i,l} n^{}_{i,l}+\sum_{\langle i, j \rangle, l, m}t^{}_{(i,l)(j,m)} c^+_{i,l} c^{}_{j,m},
\end{equation}
where $c^+_{i,l}$ ($c_{i,l}$) are creation (annihilation) operators of electrons at the orbital $l$ centered at the
$i$th site, $n_{i,l}=c^+_{i,l} c^{}_{i,l}$ is the corresponding number operator, $\varepsilon_{i,l}$ is the onsite
energy and $t_{(i,l)(j,m)}$ the hopping term.
In this equation $\langle i, j \rangle$ comprises the sum over pairs of sites for which the hopping terms are not negligible:
The summation is performed over sites inside rectangular regions like the ones in Fig.~\ref{fig:geometry}(a).
For $W=$ 1, 2 and 3 we take up to 2, 5 and 8 neighbors, respectively.
The parameters were calculated within the LCDO scheme.
In order to improve the reliability of the electronic calculations at smaler interdornor distances, we extend the treatment presented in Ref.~\cite{*[{}] [{ for details on the LCDO formalism see Suppl. Mat.}] Dusko2018} by including multi-orbitals and three-center corrections due to neighboring cores in the hopping energies.
A detailed presentation is found in the Appendix.
These developments allow us to accurately address the nanoribbon model ($W\geq2$ sites) placement parameters ($R_L$ and $R_W$) of the order of $3$nm.
We keep the isotropic approximation.

The model system we study consists of a central region, corresponding to the disordered Si:P nanostructure
coupled to leads in thermal and chemical equilibrium with electronic reservoirs, see Fig.~\ref{fig:geometry}(b).
The leads are semi-infinite, translational invariant and define the electronic bands density of states coupled
to the system of interest \cite{Lewenkopf2013}.
In addition, we investigate the effect of a uniform back gate potential,
 and study its applicability to control the nanostructure
transport properties.
The gate potential $V_G$ is included in the model as a correction to the onsite energy, namely,
$\varepsilon_{i,l}\left(V_G\right)= \varepsilon_{i,l}\left(0\right)+U_G$.
Here $U_G$ is the shift in the electronic states energy and $\varepsilon_{i,l}\left(0\right)$ is the unbiased onsite energy calculated within the LCDO scheme.
The energy gained by the electron is $U_G=\eta e V_G\propto-V_G$, where $V_G$ is the gate potential, $e$ is the electron charge, and $\eta$ is a sample-dependent constant incorporating the Si dielectric screening, geometry and the capacitive coupling of the donor electron with other leads in the system.
We expect that some trends for a lateral gate potential as the one present in Ref.\,\cite{Weber2014, Shamim2016} can be inferred by comparing different nanoribbon widths, as a confining lateral potential decreases the effective $W$.

We study the impact of positional disorder in such systems using a Gaussian disorder model.
The disorder is quantified by two parameters, namely, a cutoff radius $\delta$ around a target substitutional site
and the position standard deviation $\sigma_d$.
For simplicity, we choose $\delta=0.4$\,nm, in which case each donor can be placed at 5 different Si sites.
The degree of disorder is controlled by $\sigma_d$.
Figure ~\ref{fig:geometry}(c) gives the dependence of the distribution of the implanted ion positions
on $\sigma_d$.
The main panel shows the probability distribution $P$ of an implanted donor to occupy the aimed position
as a function of $\sigma_d$ and the inset gives a graphical representation of the disorder cutoff radius
$\delta$.
The $\sigma_d$ values considered in this work, namely,  $\sigma_d=0.1$\,nm and  $\sigma_d=0.2$\,nm are indicated
by the blue squares.
These values are within state-of-the-art precision of STM atomic placement
techniques \cite{Weber2012, Zwanenburg2013, Salfi2016}.

We calculate the nanostructure linear conductance using the Landauer-B\"uttiker formula\cite{Datta1997},
\begin{align}
\mathcal G_{AB} = \frac{2e^2}{h}\int_{-\infty}^{\infty} dE \left(-\frac{\partial f}{\partial E} \right) T_{AB}(E),
\label{eq:conductance}
\end{align}
given in terms of the Fermi-Dirac distribution function $f(E) = [1 + e^{(E-\mu)/k_B T}]^{-1}$ and the electronic transmission
$T_{AB} (E)={\rm tr} \left[ \mathbf \Gamma_B(E) \mathbf G^r(E) \mathbf \Gamma_A(E) \mathbf G^a(E) \right]$\cite{Meir1992}.
In Eq.~\eqref{eq:conductance}, $\mathbf G^r\left(\mathbf G^a\right)$ is the retarded (advanced) Green's function of the complete
system (nanoribbon and leads), which we compute using the recursive Green's function approach, implemented as in
Refs.~\cite{Lewenkopf2013, Ridolfi2017, Lima2018}.
The $n$th line or decay width, matrix elements $\mathbf{\Gamma}_n=i [ \mathbf{\Sigma}_n^r - \left( \mathbf{\Sigma}_n^r\right)^\dagger ]$
are obtained from the embedding self-energy $\mathbf{\Sigma}_n^r=\mathbf{V}_n^\dagger \mathbf{G}^r_n \mathbf{V}_n$, where
$\mathbf V_n$ contains the coupling matrix elements of the sample with the $n$th lead, while $\mathbf G^r_n$ is the contact Green's
function.
There are several ways to calculate the latter \cite{Sancho1985, MacKinnon1985, Rocha2006, Wimmer2009thesis}, we compute
$\mathbf G^r_n$ by a standard decimation procedure based on renormalization-group ideas \cite{daSilva1981, Robbins1983}.

We cast the nanostructures transport properties in terms of the  localization length $\xi$, formally defined by the wave function
asymptotic behavior, $\Psi(x)\propto \exp (-|x|/\xi)$.
In this work, we infer the localization length by the analysis of the conductance at zero temperature.

\section{Transport and Placement}
\label{sec:transport_and_placement}

Si:P nanostructures are multi-path systems due to their multi-orbital nature.
The hopping term in this multi-orbital framework plays an extremely non-trivial role,
opening and closing channels depending on the system parameters.
To improve the understanding of such system towards applications in nanodevices control,
we investigate how the disorder and placement parameters affect conductance and localization.

\begin{figure}[!htbp]
		\includegraphics[clip,width=\columnwidth]{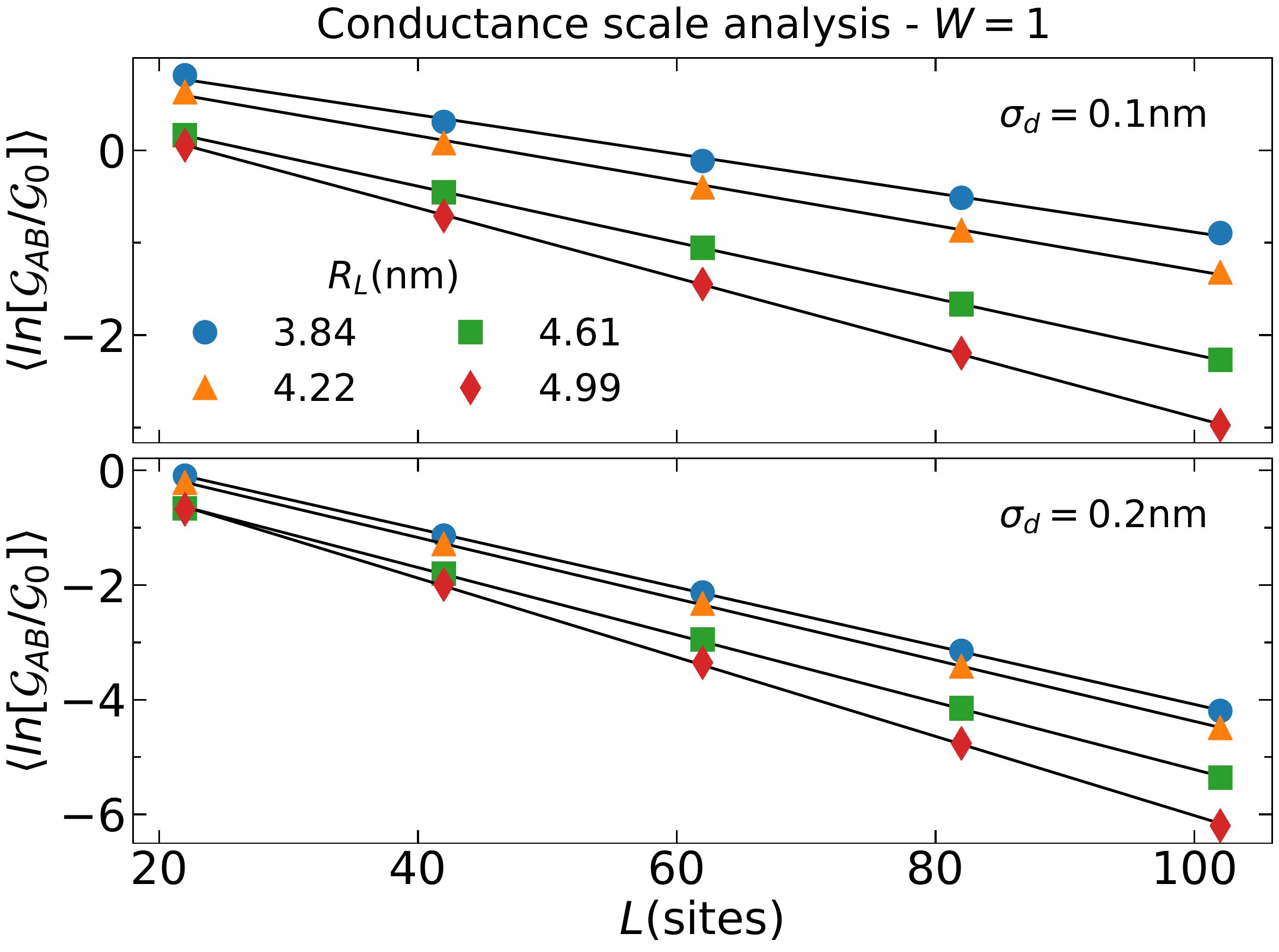}
		\caption{\label{fig:cp_conductance_vs_length}
Conductance $\mathcal{G}$ (in units of $\mathcal{G}_0=2e^2/h$) for disordered nanochains ($W=1$) as a function
of the system length $L$ (in units of $R_L$ or sites) for a few representative target interdonor distances $R_L$.
for (a)  $\sigma_d=0.1$ nm  and (b) $\sigma_d=0.2$nm.
The results correspond to an average over 10$^{4}$ disorder realizations.
In all cases the standard deviations are smaller than the markers.}
\end{figure}

According to the localization theory in disordered systems \cite{Sheng2006}, the conductance
is expected to decrease exponentially with the ratio between the sample length $L$ and the localization length $\xi$.
Hence, we extract $\xi$ from the relation $\langle \ln \mathcal{G}_{AB}(L) \rangle \propto -L / \xi$,
where $\langle \ldots \rangle$ is an ensemble average (here typically over $10^3 \cdots 10^4$ realizations).
Figure~\ref{fig:cp_conductance_vs_length} shows few representative examples of $\langle \ln \mathcal{G}_{AB} \rangle$
versus $L$ and the corresponding linear fit that gives $\xi$.

\begin{figure}[h!]
		\includegraphics[clip,width=\columnwidth]{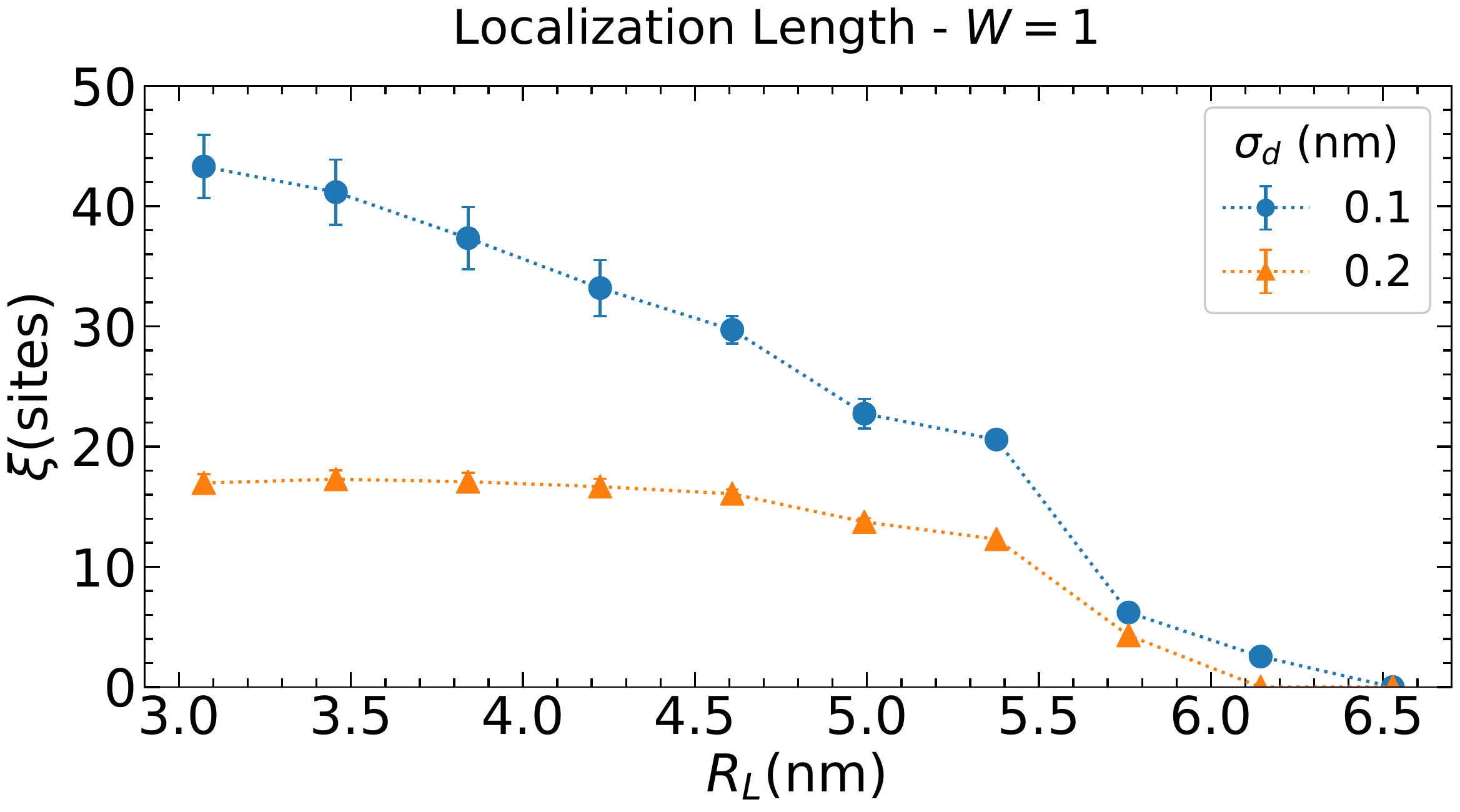}
		\caption{\label{fig:loc_length_w_1-V_g=0mV}
Localization length $\xi$ (in units of $R_L$ or sites) as a function of the interdonor target separation $R_L$ for
nanochains ($W=1$) and disorder intensities $\sigma_d=0.1$ and $0.2$ nm.
}
\end{figure}

In Fig.~\ref{fig:loc_length_w_1-V_g=0mV} we present the localization length for $W=1$ (nanochains) behavior with $R_L$ for two levels of disorder.
As expected, increasing $R_L$ or the disorder level lowers $\xi$.
Note that for small $R_L$ we observe an enhanced sensitivity of $\xi$ with $\sigma_d$.
For these two disorder levels, $\xi$ shows an abrupt fall around $R_L=5.7$\,nm.
\begin{figure*}[!htbp]
		\includegraphics[width=0.9\textwidth]{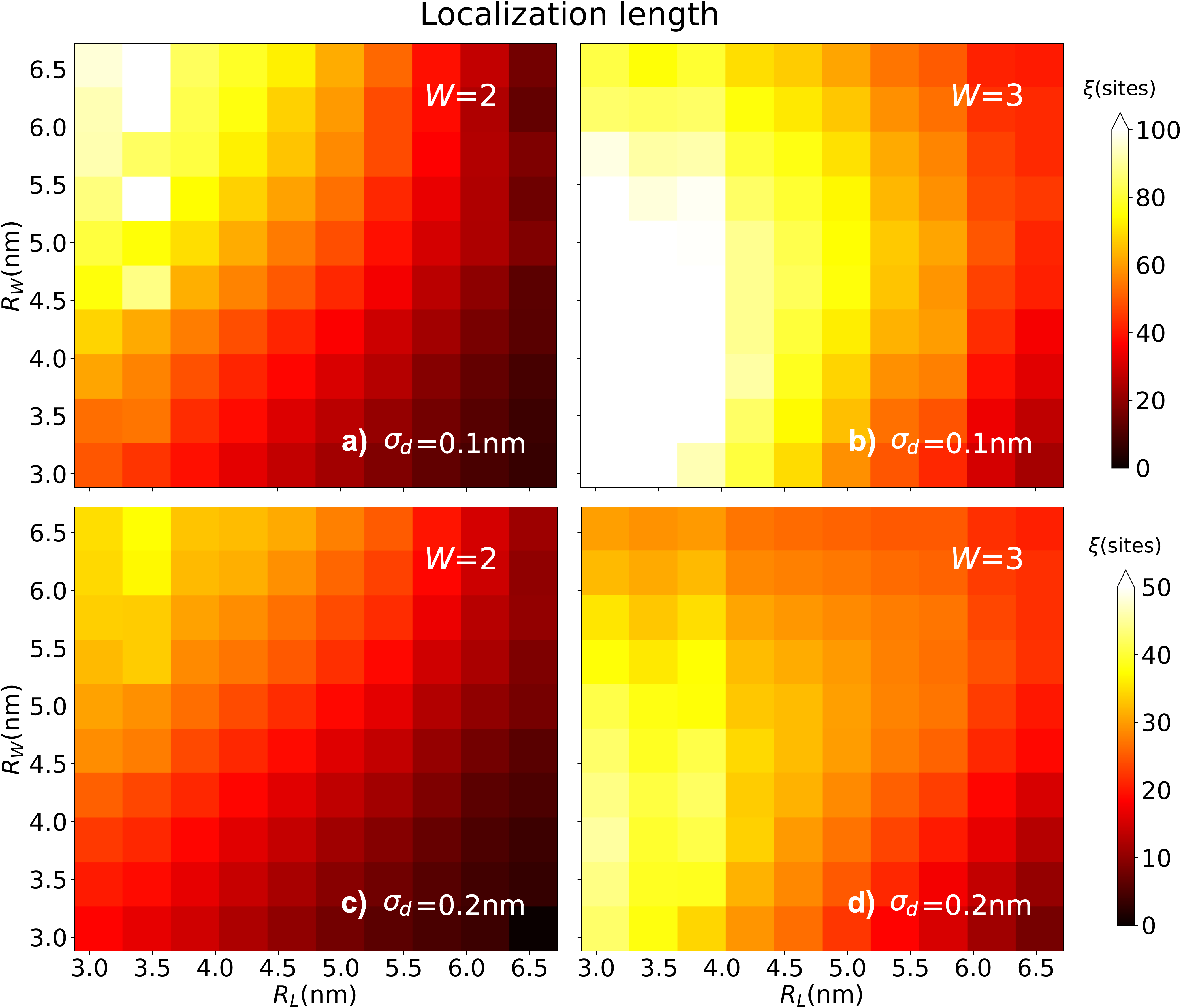}
		\caption{\label{fig:loc_length_w_2_and_3} Localization length $\xi$ for nanoribbons of $W=2$ and $3$ as a function of the
		interdonor target separation $R_L$ and $R_W$. Graphs with same $\sigma_d$ present same colorbar.
(a) $\sigma_{d}=0.1$nm and $W=2$,
(b) $\sigma_{d}=0.1$nm and $W=3$,
(c) $\sigma_{d}=0.2$nm and $W=2$ and
(d) $\sigma_{d}=0.2$nm and $W=3$.
}
\end{figure*}

In order to represent the combined effect of the geometric parameters $R_L$ and $R_W$ in the transport trends of our system, we calculate the localization length $\xi \left( R_L, R_W \right)$ for $3.0$\,nm $\lesssim \left[R_L, R_W \right]\lesssim$ $6.5$\,nm and, for each pair of parameters, $\xi$ is represented by the given color code.
In Fig.~\ref{fig:loc_length_w_2_and_3}(a) and (b) we present plots for disorder $\sigma_d=0.1$\,nm and in Fig.~\ref{fig:loc_length_w_2_and_3}(c) and (d) $\sigma_d=0.2$\,nm.
The frames on the left refer to $W=2$ and on the right to $W=3$.
The results suggest a metal-insulator phase diagram with a very similar overall behavior for both disorder intensities presented.
In Fig.~\ref{fig:loc_length_w_2_and_3} (a) and (c) our simulations reveal a relatively small region in the investigated parameter space with non-monotonic behavior, roughly $R_W\gtrsim 4.5$nm and $R_L\lesssim 5.0$nm.
In particular, $\xi$ is peaked at $R_L\approx 3.5$nm and $R_W\approx$ $5.4$, $6.1$ and $6.5$nm.
Outside this non-trivial region, by increasing $R_L$ or decreasing $R_W$ the electronic states tend to become more localized.
In Fig.~\ref{fig:loc_length_w_2_and_3} (b) and (d) we find an overall increase of $\xi$ and a wider region with
non-trivial extended states, corresponding to the parameter range defined by $R_W\alt 6.0$nm and $R_L\lesssim 4.2$nm.
In particular, $\xi$ shows peaks for $R_L\approx 3.5$nm and $R_W\approx$ $5.4$, $6.1$ and $6.5$nm.
Out of this non-trivial region, increasing $R_L$ or decreasing $R_W$ favors localization.

Comparing $W=1$, $2$ and $3$ we observe an overall increasing in localization length with the system width,
consistent with the increasing in the maximum number of transport channels, respectively $6$, $12$ and $18$.
The sensitivity of $\xi$ on the disorder intensity seems to become stronger for larger values of $W$.

\section{Tuning Localization Length}
\label{sec:loc_length}

The non-monotonic behavior of the localization length with the lattice geometry, namely $R_L$ and $R_W$,
suggests that one can tune it, and hence control the system's conductance ${\cal G}$ by a suitable external handle.
In what follows we show that a back gate potential, as described in Sec.~\ref{sec:models_and_methods}, is capable to
dramatically modify the transport properties of disordered Si:P nanowires.
We recall that for electrons, $U_G\propto-V_G$.
\begin{figure}[!htbp]
		\includegraphics[clip,width=0.9\columnwidth]{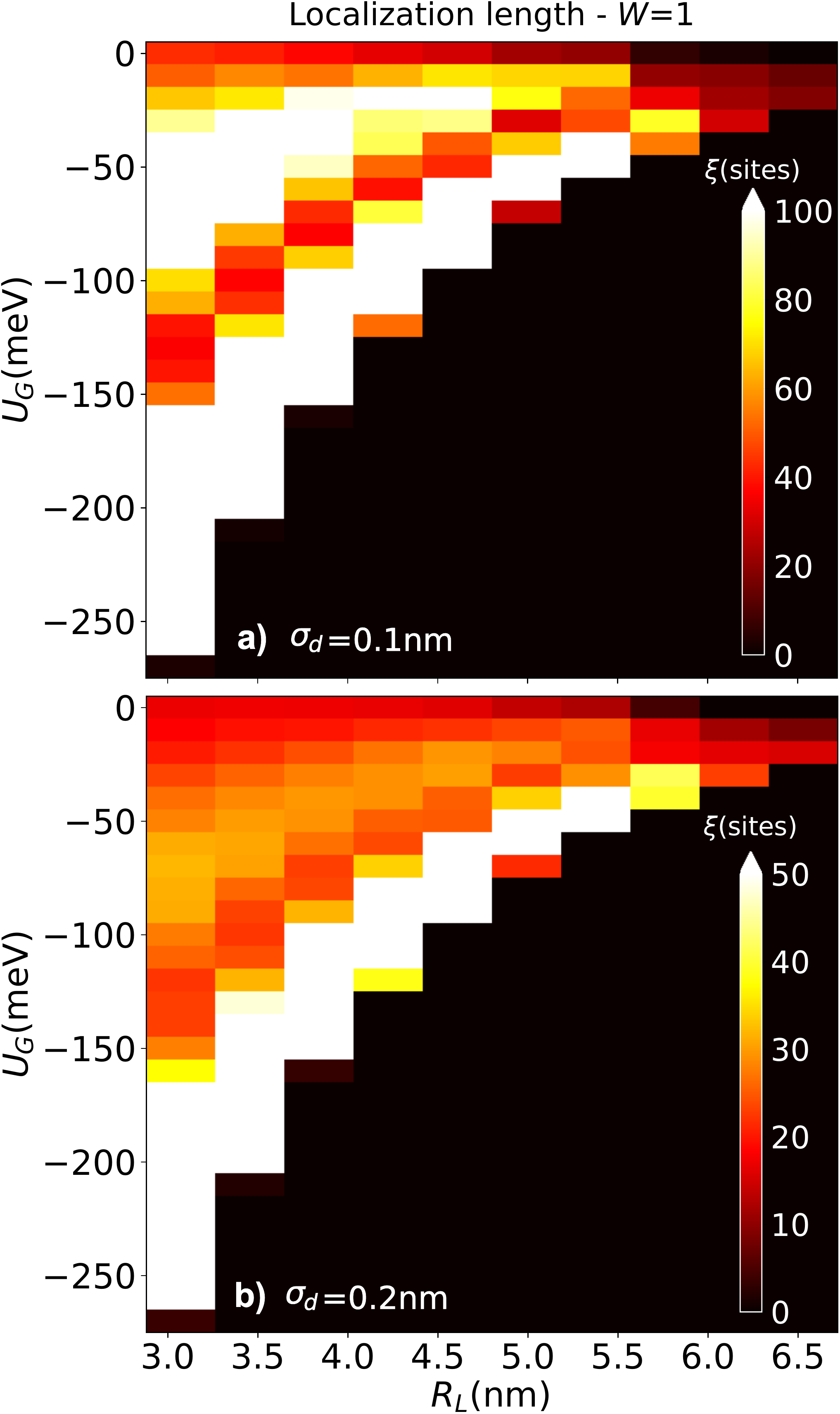}
		\caption{\label{fig:cp_loc_length_w_1_gate_pot}
Localization length $\xi$ for nanochains ($W=1$) as a function of the gate energy $U_G$ and the
interdonor target separation $R_L$ for (a) $\sigma_{d}=0.1$nm and (b) $\sigma_{d}=0.2$nm.
}
\end{figure}

In order to get some insight on the gate control over localization lengths in nanoribbons, we start with the nanowire case, $W=1$.
Results for $\xi$ under a gate bias from 0 down to $-250$\,meV are presented in Fig.~\ref{fig:cp_loc_length_w_1_gate_pot} for two degrees of disorder. For a fixed interdonor distance, according with the smaller(larger) degree of disorder $\xi$ oscillates in a larger(smaller) range in the graph truncated to $100$($50$)\,nm.
Given that the P donor in Si lower energy levels are $45$\,meV below the bottom of the Si conduction band edges, applying a bias of $U_G=45$meV would ionize the donors completely inside the active (sample) region.
A wider range of control is provided for negative values of $U_G$ which increases separation of the P electrons levels to the Si conduction band edge, thus remaining operational for the wide $U_G$ range shown in the figures.
Therefore we restrict our results to $U_G<0$ $\left(V_G>0\right)$ .
This effect can be explained as follows: $V_G$ rigidly shifts the nanowire energy spectrum.
Hence, $V_G$ drives localized and extended states, as well as small and large density of states
of the disordered system across the Fermi energy fixed by the contacts.
The parameter range for a conducting behavior ($\xi/L \agt 1$)
 shrinks for increasing values of $R_L$, consistent with the drop in the mean value of the hopping matrix elements.
An extensive analysis (not shown here) suggests a similar $\xi$ behavior with $V_G$ for different
disorder intensities.

\begin{figure*}[!htbp]
		\includegraphics[clip,width=0.9\textwidth]{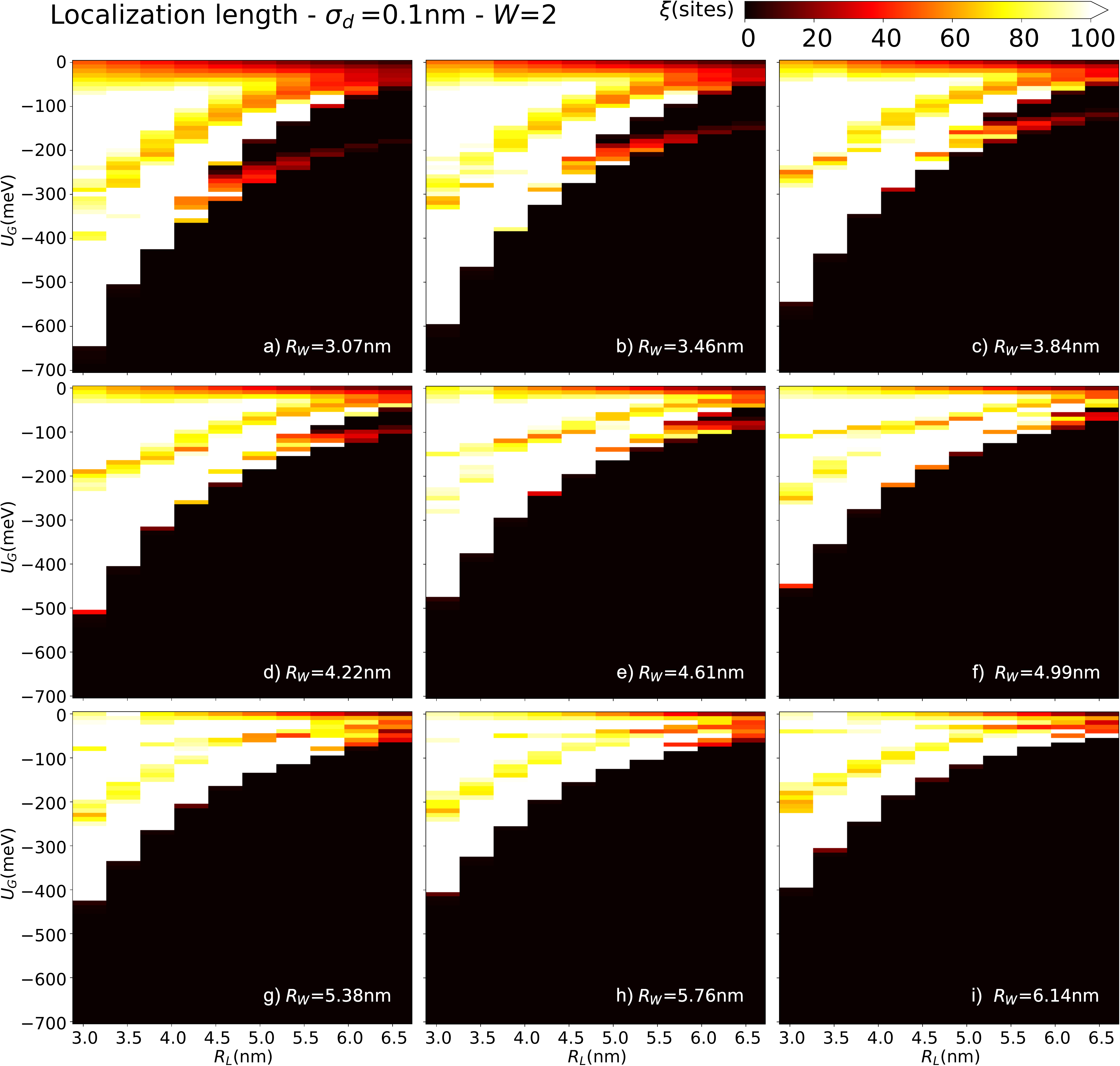}
		\caption{\label{fig:cp_loc_length_w_2_SD_01_gate_pot}
Localization length $\xi$ as a function of $V_G$ and $R_L$ ($R_W$ fixed) for nanoribbons of $W=2$ and $\sigma_d=0.1$nm.
(a) - (j) correspond to different values of $R_W$.
}
\end{figure*}

The case $W=2$ and $\sigma_d=0.1$nm is illustrated in Fig~\ref{fig:cp_loc_length_w_2_SD_01_gate_pot}.
The simulations indicate an overall increase of $\xi$ as a function of $U_G$ followed by an oscillatory pattern.
As in the $W=1$ case we observe that the $U_G$ range corresponding to conducting behavior shrinks with $R_L$.
In addition, a similar feature can be observed for  increasing values of $R_W$.
By varying $U_G$ for different $R_W$ values, we find the formation of a gap -- a region of negligible values of $\xi$ -- followed by
a ``reactivation" in the localization phase diagram for larger values of $R_L$.
This gap is highlighted in Fig.~\ref{fig:cp_loc_length_w_2_SD_01_gate_pot}(a)-(c) where the threshold $R_L$ values are $4.6$nm,
$5$nm and $5.4$nm, respectively.
The gap $R_L$ threshold value continues to increase monotonically along Fig.~\ref{fig:cp_loc_length_w_2_SD_01_gate_pot}(d)-(f).
In the last 3 panels [Fig.~\ref{fig:cp_loc_length_w_2_SD_01_gate_pot}(g)-(i)], the gap closes resembling the signature of the $W=1$ case.
Although it is reasonable to recover a phase diagram similar to $W=1$ case while increasing $R_W$, we observe an enhancement in
the overall localization length values and $V_G$ range leading to conducting behavior.

\begin{figure*}[!htbp]
		\includegraphics[clip,width=0.9\textwidth]{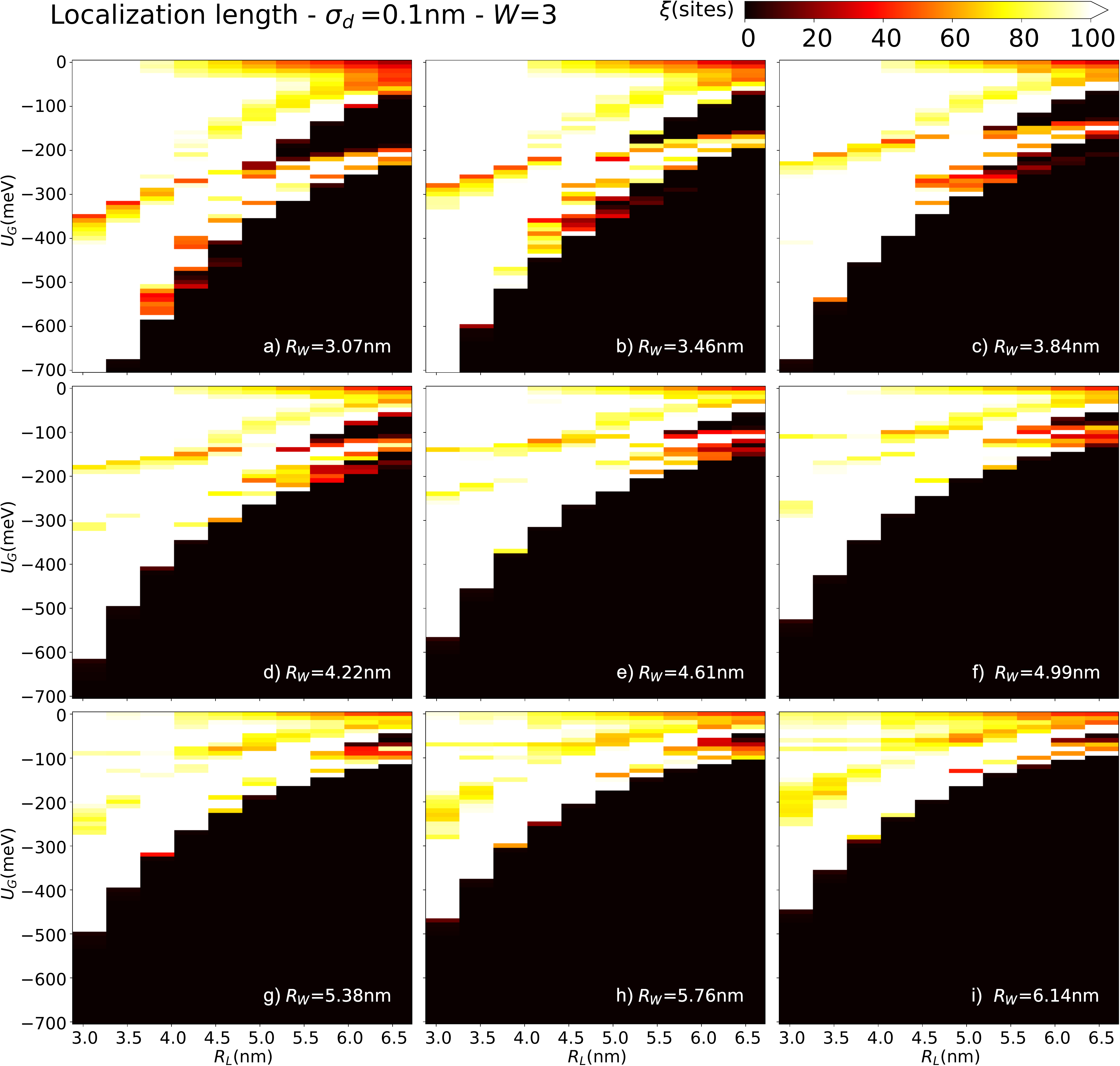}
		\caption{\label{fig:cp_loc_length_w_3_SD_01_gate_pot}
Localization length $\xi$ as a function of $V_G$ and $R_L$ ($R_W$ fixed) for nanoribbons of $W=3$ and $\sigma_d=0.1$nm.
(a) - (j) correspond to different values of $R_W$.
}
\end{figure*}

The wider ribbon case, $W=3$, is given in fig.~\ref{fig:cp_loc_length_w_3_SD_01_gate_pot}.
As in $W=1$ and $2$ cases, one observes an increase in $\xi$ with $U_G$ followed by an oscillatory pattern and that the $U_G$ range leading to conducting behavior shrinks with $R_L$ and $R_W$ interdonor distances.
For smaller $R_W$ values, Fig.~\ref{fig:cp_loc_length_w_3_SD_01_gate_pot}(a)-(c) shows a larger gap than in the $W=2$ case and the
opening of a second gap.
Throughout Fig.~\ref{fig:cp_loc_length_w_3_SD_01_gate_pot}(d)-(f) we observe that this second gap is short-lived comparing the first one.
In summary, we find that both $\xi$ and $U_G$ range leading to conducting behavior are overall larger than in the $W=1$ and $2$ cases.

We have also performed calculations for $\sigma_d=0.2$\,nm, for both $W=2$ and $3$, not shown since all properties follow the trends identified in the previous cases.
\section{Conductance control}\label{sec:conductance}

In this section, we investigate the use of a gate potential $V_G$ $\left(U_G\propto-V_G\right)$ as an external control of conductance
$\mathcal{G}_{AB}$ for Si:P nanostructures.
We set $L=60$ sites for the purpose of investigating a nanoswitch implementation in a length comparable to some experimental realizations\cite{Weber2012, Weber2014, Shamim2016} of higher P density.

\begin{figure}[!htbp]
		\includegraphics[clip,width=0.9\columnwidth]{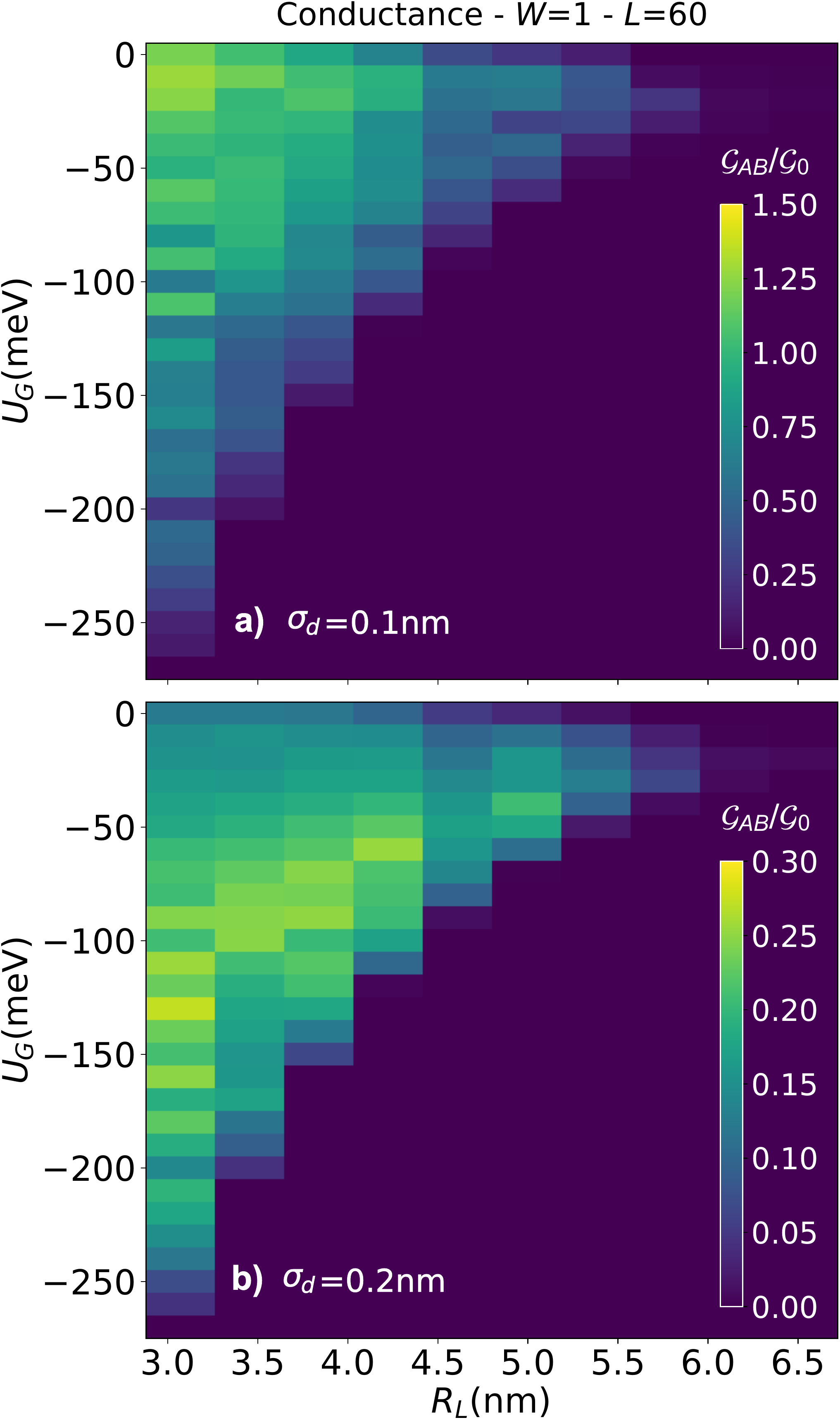}
		\caption{\label{fig:cp_conductance_w_1_gate_pot}
Conductance $\mathcal{G}_{AB}$ (in units of $\mathcal{G}_{0}=2e^2/h$) as a function of a gate potential $V_G$ and interdonor
target separation $R_L$ for nanochains ($W=1$), $L=60$ sites and (a) $\sigma_{d}=0.1$nm (b) $\sigma_{d}=0.2$nm.
}
\end{figure}

Figure ~\ref{fig:cp_conductance_w_1_gate_pot} shows the average conductance  $\mathcal{G}_{AB}$
as a function of $U_G$ and $R_L$ for nanochains ($W=1$),
The results show oscillations in $\mathcal{G}_{AB}$ as a function of both $R_L$ and $V_G$.
A minimum in $\mathcal{G}_{AB}$ occurs around $R_L\approx4.6$ nm, which should be avoided in practical implementations of the system as a nanoswitch.
Oscillations due to $U_G$ stand out for smaller $R_L$ values. In line with the localization length analysis,
an increase of $R_L$ causes the range of $U_G$ values corresponding to a conducting behavior to shrink.
For $R_L\approx3.1$ nm, introducing a gate potential, we observe an increase in $\mathcal{G}_{AB}$ of
approximately $50\%$ and $100\%$ for $\sigma_d=0.1$nm and $0.2$nm, respectively.

\begin{figure*}[!htbp]
		\includegraphics[clip,width=0.9\textwidth]{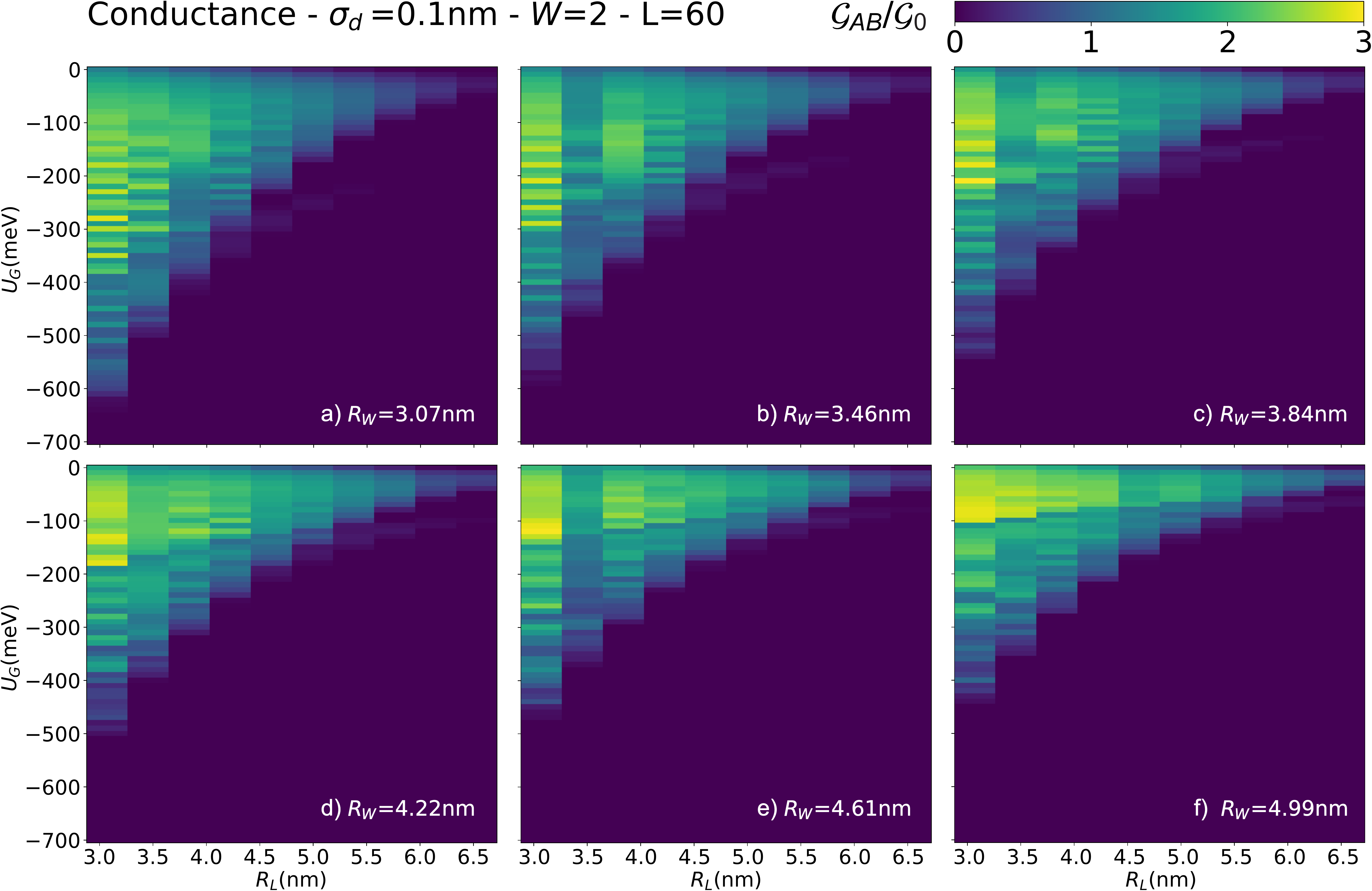}
		\caption{\label{fig:cp_conductance_w_2_SD_01_gate_pot}
Conductance $\mathcal{G}_{AB}$ (in units of $\mathcal{G}_{0}=\frac{2e^2}{h}$) for disordered nanoribbons of $W=2$ and $L=60$ sites with $\sigma_d=0.1$nm as a function of a gate potential $V_G$ and interdonor target separations $R_L$ and $R_W$.
(a) - (f) correspond to different values of $R_W$.
}
\end{figure*}

The conductance for $W=2$ sites nanoribbons and $\sigma_d=0.1$nm results, presented in Fig.~\ref{fig:cp_conductance_w_2_SD_01_gate_pot}, show a rapidly oscillatory behavior as a function of $U_G$ for small $R_W$ values, see Fig.~\ref{fig:cp_conductance_w_2_SD_01_gate_pot}(a)-(b).
For larger $R_W$ values however [see Fig.~\ref{fig:cp_conductance_w_2_SD_01_gate_pot}(e)-(f)] the oscillations are strongly damped for small $U_G$.
In all cases, it is possible to observe a $U_G$ transition edge between larger and smaller $\mathcal{G}_{AB}$ values regimes.
There is also a minimum in $\mathcal{G}_{AB}$ around $R_L\approx3.5$nm, the feature is more pronounced in the cases shown in Fig.~\ref{fig:cp_conductance_w_2_SD_01_gate_pot}(b)-(e).
We observe a very subtle gap opening in $\mathcal{G}_{AB}$ while increasing $R_L$.
The $R_L$ value corresponding to this opening increases with $R_W$.
In Fig.~\ref{fig:cp_conductance_w_2_SD_01_gate_pot}(a), (c) and (e) the corresponding gapping opening value
is $R_L\approx4.2$, $5.0$ and $5.4$\,nm, respectively.
 As in the $W=1$ case, introducing a gate potential induces an increase of approximately 50\% in $\mathcal{G}_{AB}$.

\begin{figure*}[!htbp]
		\includegraphics[clip,width=\textwidth]{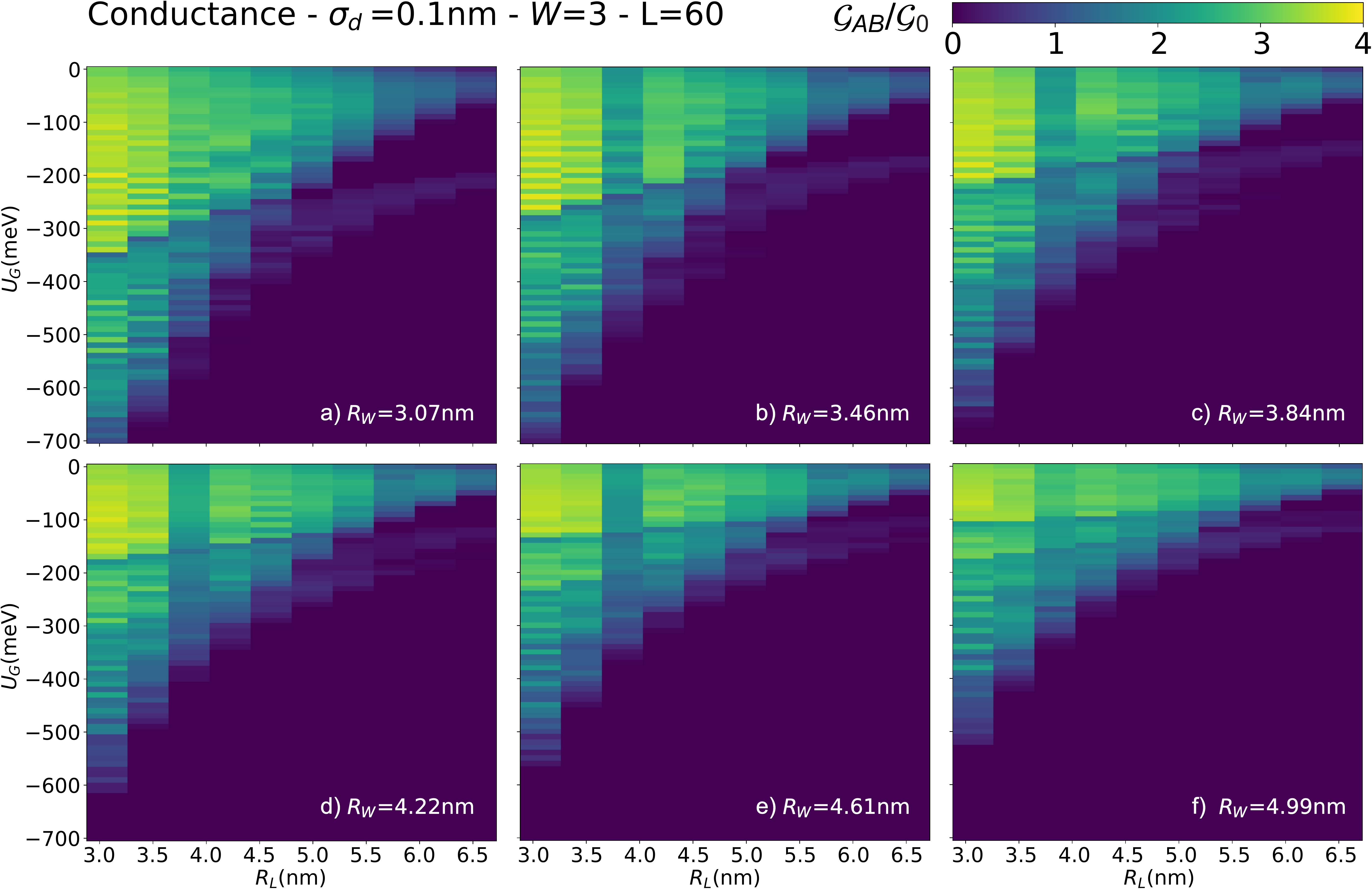}
		\caption{\label{fig:cp_conductance_w_3_SD_01_gate_pot}
Conductance $\mathcal{G}_{AB}$ (in units of $\mathcal{G}_{0}=\frac{2e^2}{h}$) for nanoribbons of $W=3$ and $L=60$ sites with $\sigma_d=0.1$nm as a function of a gate potential $V_G$ and interdonor target separations $R_L$ and $R_W$.
(a) - (f) correspond to different values of $R_W$.
}
\end{figure*}

The results for nanoribbons of $W=3$ sites are presented in Fig.~\ref{fig:cp_conductance_w_3_SD_01_gate_pot}.
Some similarities with $W=2$ case can be observed: Rapidly oscillating $\mathcal{G}_{AB}$ spectrum with a clear change in overall behavior in a given transition edge, for an example see Fig.~\ref{fig:cp_conductance_w_3_SD_01_gate_pot}(a) at $R_L\approx3.1$\,nm and $U_G=-350$\,meV.
In contrast with the $W=2$ case we observe two gap openings and an overall minimum in $\mathcal{G}_{AB}$ values around $R_L\approx3.9$\,nm.
The first gap can be observed in Fig.~\ref{fig:cp_conductance_w_3_SD_01_gate_pot}(a), (c) and (e) for $R_L\approx5.0$, $5.4$ and $5.9$\,nm, respectively.
The second gap is more subtle but can be observed in Fig.~\ref{fig:cp_conductance_w_3_SD_01_gate_pot}(d) for $R_L\approx5.9$\,nm, for example.

In summary, by considering nanostructures with increasing width, $W=1$, $2$ and $3$, we observe a
corresponding increase in:
(i) the overall $\mathcal{G}_{AB}$ values;
(ii) the window of $V_G$ values leading to a conducting behavior;
and (iii) in the number of gap openings.
We also observe a change in the $R_L$ value corresponding to an overall minimum in $\mathcal{G}_{AB}$.
We also find that the overall behavior of the localization length on the lattice parameters does not depend on
the disorder strength. This can be explicitly seen for the $W=1$ case by comparing the simulations for
$\sigma_d=0.2$\,nm and $0.1$\,nm.

Finally, let us stress the sharp $V_G$ driven metal-insulator transition appearing for any given choice of $R_L$ in all
cases we analyze in this work.
This remarkable feature strongly suggest that Si:P nanostructures can act as switches by properly tuning the gate potential.

\section{Discussions and Conclusions}
\label{sec:conclusions}

In this work we extended the LCDO formalism\,\cite{Dusko2016,Dusko2018} to include Gaussian disorder, a multi-orbital description and technical improvements, specified in the Appendix, which results in a more realistic description of P nanochains and to access nanoribbons of arbitrary widths.
Our simulations treat the problem considering realistic system sizes and disorder.
We also have put forward a proposal for an external control of transport properties such as localization length and conductance suggesting a new path of investigations for future experimental implementations.

We have found nontrivial features of the electronic transport properties due to system fabrication specifications still remaining robust against disorder.
Specific values of placement parameters and nanostructure width provide optimized localization length, favoring high conductance.
Our calculations indicate that a similar behavior is expected for different disorder levels.

We further analyze the effects of an external back gate potential $V_G$ to localization
length and conductance.
Properly tuning $V_G$ one can control localization lengths, allowing donor nanowires to keep current-carrying wave functions even for relatively long samples, serving as efficient connectors among nanodevices parts.
In addition, it is possible to increase the nanostructure conductance, or decrease it by using this external potential, which suggests the use of such structures as nanoswitches.
Both connectors and switches provide state-of-the-art resources contributing to nanodevices technology development.

\acknowledgments
The authors acknowledge the financial support of the Brazilian funding agencies
CL CNPq (grant  308801/2015-6); 
BK CNPq (grant 304869/2014-7) and FAPERJ (grant E-26/202.767/2018).
This study was also financed in part by the Coordena\c{c}\~ao de Aperfei\c{c}oamento de Pessoal
de N\'{\i}vel Superior - Brazil (CAPES) - Finance Code 001.

\appendix*
\section{Microscopic model - Technical details}
\subsection{Linear Combination of Dopant Orbitals (LCDO)}

Following the well established Kohn and Luttinger prescription~\cite{Luttinger1955, Kohn1955, Kohn1957} for shallow donors in Si, we consider a basis of six donor orbitals per site, corresponding to the six minima in Si conduction band.
Valley orbit coupling, included by first order perturbation theory for degenerate states~\cite{Koiller2002,Saraiva2015}, renders donor orbitals as superpositions of pure valley states obtained by the effective mass approach:
\begin{equation}
\label{eq:wave_function}
    \Psi_i^{l} ({\bf r}) ={\frac{1}{N_l}} \sum_{\mu=1}^6 a^l_\mu F_\mu ({\bf r}- {\bf R}_i) \phi_\mu ({\bf r}- {\bf R}_i),
\end{equation}
where $l$ refer to the donor $i$ orbitals pinned to the donor coordinates ${\bf R}_i$.
The constants $N_l$ and $a^l_\mu$ stand for the normalization and valley population (presented in
Table~\ref{tab:donor_coupled_states}), $ F_{\mu} ({\bf r})=F (r) =(\pi a^{*3})^{-1/2} \, e^{-r/a^*}$ is
for simplicity approximated as an isotropic hydrogen-like envelope function, with a species dependent effective Bohr radius  $a^*$
 ($1.106$\,nm for Si:P) and $\phi_\mu ({\bf r})=e^{i{k_\mu}\cdot{\bf r}} u_\mu\left({\bf r} \right)$ are the Bloch
 functions of the 6 Si conduction band degenerate minima ($\mu= 1, \cdots,  6$).
The latter are located along the equivalent directions $\pm x$, $\pm y$, $\pm z$ at $\left| {\bf k}_\mu \right|=
k_0=0.85 (2 \pi /a_{\rm Si})$, where $a_{\rm Si}$ is the conventionally called Si lattice parameter
\cite{Madelung2012}.
The effective Bohr radius is obtained by incorporating screening effects due to the Si host charge carriers
in the donor singular potential.

\begin{table}[!htbp]
\caption{\label{tab:donor_coupled_states}
 Valley population $a_\mu^l$, normalization constant $N_l$, and P$^0$ donor energy $E_l$ for the 6 donor orbitals $l$. }
\begin{ruledtabular}
\begin{tabular}{ccccccccc}
 $l$        & $a_{x}^l$ & $a_{-x}^l$& $a_{y}^l$ & $a_{-y}^l$& $a_{z}^l$ & $a_{-z}^l$& $N_l$         & $E_l$(meV)\\ \\[-8pt]
\hline \\[-8pt]
A$_{1}$    & 1         & 1         & 1         & 1         & 1         & 1         & $\sqrt{6}$    &-45.58 \\ \hline \\[-8pt]
 T$_{2}^z$  & 0         & 0         & 0         & 0         & 1         & -1        & $\sqrt{2}$    &\\
T$_{2}^y$   & 0         & 0         & 1         & -1        & 0         & 0         & $\sqrt{2}$    &-33.90\\
T$_{2}^x$   & 1         & -1        & 0         & 0         & 0         & 0         & $\sqrt{2}$    & \\ \\[-8pt]
\hline \\[-8pt]
 E$^{xy}$   & 1         & 1         & -1        & -1        & 0         & 0         & $2$           &-32.60\\
 E$^{z}$    & 1         & 1         & 1         & 1         & -2        & -2        & $\sqrt{12}$   & \\
\end{tabular}
\end{ruledtabular}
\end{table}

Screening effects are included through a potential that interpolates the expected
behavior for large and small values of $r$, namely,
\begin{equation}
  V(r)=-\frac{ e^2}{4 \pi r}\left[\frac{1}{\epsilon_{\rm Si}} + \left( \frac{1}{\epsilon_{0}} - \frac{1}{\epsilon_{\rm Si}} \right)e^{-{r}/{r^{*}}} \right],
\end{equation}
the screening length $r^{*}$ defines the transition between a bare $V(r\rightarrow0)={-e^2}/{4 \pi \epsilon_{0} r}$ and
a screened $V(r\rightarrow\infty)={-e^2}/{4 \pi \epsilon_{\rm Si} r}$ potential.
Here $\epsilon_{0}$ and $\epsilon_{\rm Si}$ are respectively the free space and the static relative permittivities.

As in previous works \cite{Dusko2016, Dusko2018} the Hamiltonian terms are calculated by the atomistic
Hamiltonian $\hat{H}= \hat{H}_i+\hat{H}'$, where $\hat{H}_i$ is the single donor Hamiltonian and $\hat{H}'$ is
the perturbation due to neighboring donor cores.
We project the donor orbital to this atomistic Hamiltonian to extract the onsite and hopping terms.
For the onsite term we obtain,
\begin{eqnarray}
\label{eq:hopping_parameter}
 \varepsilon_{i,l} = \langle i |\hat{H_i}|i\rangle +\langle i|\hat{H}'|i\rangle
 \approx  - E_l+\sum_k \langle i|\hat{V}_k|i\rangle,
 \end{eqnarray}
where $E_l$ is the single donor level energy given in Table~\ref{tab:donor_coupled_states}, which contains
valley-orbit corrections.

Similarly the hopping reads,
  \begin{subequations}
  \begin{eqnarray}
\label{eq:hopping_parameter}
 t_{(i,l)(j,m)}         &=& \langle j |\hat{H_i}|i\rangle +\langle j|\hat{H}'|i\rangle \\
                        &\approx&  - E_0 \langle j |i\rangle+\sum_k \langle j|\hat{V}_k|i\rangle
                        = \mathcal{T}_{ij} (R) \Theta^{lm} ({\bf R}) \nonumber \\
 \Theta^{lm}            &=&\frac{1}{N_l N_m}\sum_{\mu, \nu=1}^{6} a_{\mu}^l a_{\nu}^m e^{i \bf{k_{\mu} \cdot R}} \\
 \mathcal{T}_{ij} (R)   &=& E_0 \mathcal{S}_{ij}+T_{ijj}+\sum_k T_{ikj} \\
 \mathcal{S}_{ij} (R)   &=& \langle F({\bf R}{_j}) | F({\bf R}{_i})\rangle \\
 T_{ikj}                &=& \langle F({\bf R}{_j}) | V({\bf R}{_k}) | F({\bf R}{_i})\rangle,
 \end{eqnarray}
\end{subequations}
where ${\bf R}={\bf R}_j-{\bf R}_i$ is the interdonor distance, $E_0$ is the donor ground state energy,
$\Theta^{lm}$ comes from the valley interference, and $\mathcal{T}_{ij} (R)$ depends on the envelope
overlap function $\mathcal{S}_{ij}$ and on two-centers ($T_{ijj}$) and three-centers ($T_{ikj}$) envelope
function integrals.
The $T_{ijj}$ integrals have a closed analytical solution~\cite{Dusko2018},
while the $T_{ikj}$ are calculated numerically.
The $k$ labels all cores in the neighborhood of the $i$ and $j$ donors, see Fig.~\ref{fig:geometry}a.

Comparisons with experiments show that this multivalley central cell corrected dopant approximation
gives an accurate description of the single impurity spectrum\cite{Saraiva2015} and the corresponding wave
functions\cite{Saraiva2016}, as well as the two impurities spectra in ionized\cite{Gonzalez2014} and neutral
excited states\cite{Dehollain2014}.
The computationally advantage is clear: By incorporating the Si matrix explicitly in the orbitals, this approach
allows the investigation of shallow donor systems of mesoscopic dimensions, a prohibitive task for a full
atomistic approach.

\subsection{Gaussian Expansion - Three-center Integrals}

In this paper we consider hopping terms due to all neighboring cores.
Since the straightforward calculation of these three-center integrals is computationally expensive,
we write the envelope orbitals and the Coulomb potential, as a Gaussian expansion, namely
\begin{eqnarray}
  F(r)  &=&\sum_{n=1}^{N_{\rm G}} c^F_n e^{-s^F_n r^2}, \\
  V(r)  &=&-\frac{ e^2}{4 \pi r}\left[\frac{1}{\epsilon_{\rm Si}} + \left( \frac{1}{\epsilon_{0}} - \frac{1}{\epsilon_{\rm Si}} \right)
  \sum_{n=1}^{N_{\rm G}} c^V_n e^{-s^V_n r^2} \right], \qquad
\end{eqnarray}
where the coefficients $c^F_n$, $c^V_n$, $s^F_n$, and $s^V_n$ are obtained by a standard least square fit and presented in
Table~\ref{tab:Gauss_expansion}.
We find that by taking $N_G=13$ Gaussian terms, the expansions agree within $10^{-8}$ accuracy for all values of $r$ where
the target function satisfies $f(r) \agt10^{-20}$.

\begin{table}[!htbp]
\caption{\label{tab:Gauss_expansion} Gaussian expansion coefficients for the envelope function $F(r)$ and
exponential in the screened Coulomb potential $V(r)$.}
\begin{tabular}{cccc}
    \hline\hline \\[-8pt]
    \multicolumn{2}{C{4cm}}{$F(r)$} &  \multicolumn{2}{C{4cm}}{$V(r)$}  \\ \hline \\[-8pt]
     $c^F_n$    &   $ s^F_n (\textrm{nm}^{-2})$   & $c^V_n$  & $s^V_n(\textrm{nm}^{-2})$  \\ \hline \\[-8pt]
    $9.26 \times 10^{-2}$ & $4.10 \times 10^{-1}$&  $1.91 \times 10^{-1}$& $3.82 \times 10^{ 1}$ \\
    $8.56 \times 10^{-2}$ & $9.19 \times 10^{-1}$ & $1.76 \times 10^{-1}$ & $8.60 \times 10^{ 1}$ \\
    $7.31 \times 10^{-2}$ & $1.89 \times 10^{-1}$ &  $1.52 \times 10^{-1}$ & $1.76 \times 10^{ 1}$ \\
    $6.71 \times 10^{-2}$ & $2.15                       $ &  $1.38 \times 10^{-1}$ & $2.02 \times 10^{ 2}$ \\
    $4.80 \times 10^{-2}$ & $5.30                       $ &  $9.86 \times 10^{-2}$ & $4.99 \times 10^{ 2}$ \\
    $3.26 \times 10^{-2}$ & $1.38 \times 10^{ 1}$ &  $6.68 \times 10^{-2}$ & $1.31 \times 10^{ 3}$ \\
    $3.13 \times 10^{-2}$ & $8.99 \times 10^{-2}$ &  $6.53 \times 10^{-2}$ & $8.35               $ \\
    $2.12 \times 10^{-2}$ & $3.89 \times 10^{ 1}$ &  $4.34 \times 10^{-2}$ & $3.70 \times 10^{ 3}$ \\
    $1.33 \times 10^{-2}$ & $1.20 \times 10^{ 2}$ &  $2.72 \times 10^{-2}$ & $1.15 \times 10^{ 4}$ \\
    $8.03 \times 10^{-3}$ & $4.20 \times 10^{ 2}$ &  $1.63 \times 10^{-2}$ & $4.05 \times 10^{ 4}$ \\
    $4.68 \times 10^{-3}$ & $1.78 \times 10^{ 2}$ &  $9.47 \times 10^{-3}$ & $1.74 \times 10^{ 5}$ \\
    $3.84 \times 10^{-3}$ & $4.26 \times 10^{-2}$ &  $8.03 \times 10^{-3}$ & $3.96               $ \\
    $3.73 \times 10^{-3}$ & $1.54 \times 10^{ 4}$ &  $7.45 \times 10^{-3}$ & $1.53 \times 10^{ 6}$ \\
    \hline\hline
\end{tabular}
\end{table}

\subsection{Gaussian Coulomb Integrals - Product Rule}

Let us now show the main derivation steps to obtain very simple expressions
for the Gaussian integrals introduced above.
The Gaussian expansion of the Coulomb three-center integral $T_{acb}$ reads
\begin{eqnarray}
  T_{acb}   &=& \langle F({\bf R}{_b}) | V({\bf R}{_c}) | F({\bf R}{_a})\rangle \\
            &=& \sum_{m,n} c^F_m c^F_n \int_{V} d{\bf r} \, e^{-s^F_m r_b^2} e^{-s^F_n r_a^2} V({\bf r}{_c}), \nonumber
\end{eqnarray}
where $r_n=\left|{\bf r} - {\bf R}{_n} \right|$ is the relative position to donor $n$.

Let us now use the Gaussian product rule, i.e.,
\begin{eqnarray} \label{eq:product_rule}
    e^{-s^F_m r_b^2} e^{-s^F_n r_a^2}   &=& e^{-\eta_{mn} R_{ba}^2} e^{-u_{mn} r_u^2},
\end{eqnarray}
where the constants $u_{mn}=s^F_m + s^F_n$ and $\eta_{mn}={s^F_m  s^F_n}/{u_{mn}}$ are
the total and reduced exponents, while $R_{ba}=\left| {\bf R}{_a} -{\bf R}{_b} \right|$ and
$r_u={\left(s^F_m r_b + s^F_n r_a\right)}/{u_{mn}}$ are the relative and the Gaussian center
of mass positions.
Equation \eqref{eq:product_rule} expresses the product of two Gaussians in a new product where
the first term is a constant and only the second term depends on ${\bf r}$. In other words, the problem
is reduced to a two-center integral
\begin{eqnarray}
  T_{acb}&=& \sum_{m,n} c^F_m c^F_n e^{-\eta_{mn} R_{ba}^2}  \int_V  \! d{\bf r} \,e^{-u_{mn} r_u^2} V({\bf r}{_c}) .
\end{eqnarray}

When $V({\bf r})$ is a screened Coulomb potential, this two-centers integral can be decomposed in
two terms, i.e. $T_{F}\left({\bf r}{_u}, {\bf r}{_c} \right)= T_{Si}\left({\bf r}{_u}, {\bf r}{_c} \right)+T_{sc}\left({\bf r}{_u}, {\bf r}{_c} \right)$.
The Gaussian expansion of the exponential term in the $V(r)$ gives
\begin{eqnarray}
  T_{\rm Si}            &=&-\frac{ e^2}{4 \pi \epsilon_{Si}}\sum_{m,n} c^F_m c^F_n e^{-\eta_{mn} R_{ba}^2}  \mathcal{I}^{\rm Si}_{mn} \nonumber, \\
  \mathcal{I}^{\rm Si}_{mn}  &=&\int_V \! d{\bf r}  \, e^{-u_{mn} r_u^2} \frac{1}{r_c}  \nonumber, \\
  T_{sc}            &=&-\frac{ e^2}{4 \pi} \left( \frac{1}{\epsilon_{0}} - \frac{1}{\epsilon_{\rm Si}}\right) \sum_{m,n,o} c^F_m c^F_n c^V_o e^{-\eta_{mn} R_{ba}^2} \mathcal{I}^{sc}_{mno} \nonumber,\\
  \mathcal{I}^{sc}_{mno}  &=& \int_V \!d{\bf r}\, e^{-u_{mn} r_u^2} \frac{e^{-s^F_o r_c^2}}{r_c}  \nonumber.
\end{eqnarray}

The next step consists in writing $r_c^{-1}$ as  Gaussian integral, namely, $r_c^{-1}=\pi^{-1/2}\int_{-\infty}^{\infty}\! dt\, e^{-t^2 r_c^2}$.
After rearranging the integrals and applying the product rule in the $sc$ term,  one obtains
\begin{eqnarray}
    \mathcal{I}^{\rm Si}_{mn}   &=&\frac{1}{\sqrt{\pi}} \int_{-\infty}^{\infty} \!dt \int_V \!d{\bf r}\, e^{-u_{mn} r_u^2} e^{-t^2 r_c^2}   \nonumber, \\
    \mathcal{I}^{sc}_{mno}  &=&\frac{1}{\sqrt{\pi}} e^{{-\nu_{mno} R_{uc}^2}}\int_{-\infty}^{\infty} \!dt \int_V \! d{\bf r} \, e^{-v_{mno} r_v^2} e^{-t^2 r_c^2}  \nonumber,
\end{eqnarray}
where $v_{mno}=( u_{mn} + s^F_o)$, $\nu_{mno}= u_{mn}s^F_o/v_{mno}$,
$R_{uc}=| {\bf R}_c - {\bf R}_u |$ and $r_v=|{\bf r} - {\bf R}_v |$ where
${\bf R}_v=(u_{mn} {\bf R}_u+s^F_o {\bf R}_c)/v_{mno}$. Applying the product
rule, as in Eq.~\eqref{eq:product_rule}, we find
\begin{eqnarray}
    \mathcal{I}^{\rm Si}_{mn}  \!  &=& \! \frac{1}{\sqrt{\pi}} \int_{-\infty}^{\infty}\!
    dt \, e^{-\left(\frac{u_{mn} t^2}{u_{mn}+t^2}\right) R_{uc}^2 }
    \int_V \!\! d{\bf r}\, e^{-\left(u_{mn} + t^2 \right) r_p^2} \nonumber
\end{eqnarray}
where $r_p=( u_{mn} r_u + t^2 r_c) /(u_{mn} + t^2)$ and
\begin{eqnarray}
    \mathcal{I}^{sc}_{mno} =
    \frac{e^{{-\nu_{mno} R_{uc}^2}}}{\sqrt{\pi}} \! \! \int_{-\infty}^{\infty}
    \! \! \! \!dt\, e^{\! -\left(\frac{v_{mno} t^2}{v_{mno}+t^2}\right) R_{vc}^2 }
    \! \! \int_V \! \! \!d{\bf r} \, e^{\! -\left(v_{mno} + t^2 \right) r_q^2}  \nonumber
\end{eqnarray}
where $r_q= ( v_{mno} r_v + t^2 r_c) / (v_{mno} + t^2 )$ and  $R_{vc}=| {\bf R}{_c} - {\bf R}{_v} |$.
As in Eq.~\eqref{eq:product_rule}, the spatial integrals depend only in the Gaussian center of mass $r_p$ and $r_q$.

Finally, adjusting the integration limit in the remaining integrals, we obtain the simple expressions
\begin{eqnarray}
    \mathcal{I}^{\rm Si}_{mn}   &=&\frac{2}{\sqrt{\pi}} \int_{0}^{\infty} \left( \frac{\pi}{u_{mn} + t^2} \right)^{3/2}  \!\!e^{-\left(\frac{u_{mn} t^2}{u_{mn}+t^2}\right) R_{uc}^2 } \,dt \nonumber, \\
    \mathcal{I}^{sc}_{mno}  &=&\frac{2e^{{-\nu_{mno} R_{uc}^2}}}{\sqrt{\pi}} \! \int_{0}^{\infty} \!\! dt \left( \frac{\pi}{v_{mno} + t^2} \right)^{3/2}  \!\! e^{-\left(\frac{v_{mno} t^2}{v_{mno}+t^2}\right) R_{vc}^2 }  \nonumber.
\end{eqnarray}
By introducing the change of variables $q_u^2={t^2}/\left({u_{mn}+t^2}\right)$ and $q_v^2={t^2}/\left({v_{mno}+t^2}\right)$ the integrals are
conveniently written as
\begin{eqnarray}
    \mathcal{I}^{\rm Si}_{mn}   &=&{2 \pi} \int_0^1 \!dq_u\, e^{-u_{mn}R_{uc}q_u^2}={2 \pi}F_0\left[{u_{mn}}R_{uc}^2\right] \nonumber, \\
    \mathcal{I}^{sc}_{mno}  &=&{2 \pi e^{{-\nu_{mno} R_{uc}^2}}} \int_0^1 \! dq_v \,e^{-v_{mn}R_{vc}q_v^2} \\
                            &=&{2 \pi e^{{-\nu_{mno} R_{uc}^2}}}F_0\left[{v_{mn}}R_{vc}^2\right] \nonumber,
\end{eqnarray}
where $F_0$ is called zero degree Boys function \cite{Boys1950, Gill1994}.
To optimize computational resources we choose to solve the integral once, with high precision and in a range
covering small and large values, and to adjust a curve that interpolates with rapidly decaying exponentials the
expected behavior in all domain.
\begin{equation}
\label{eq:Boys_fitted_curve}
    F_0^{\rm adj}(x)=e^{-s_1 x^6}\sum_{n=0}^{6}B_n x^n+\left( 1-e^{-s_2 x^6} \right)\frac{1}{2}\sqrt{\frac{\pi}{x}}
\end{equation}
The coefficients of the fitted curve are presented in Table~\ref{tab:Boys_expansion}. For the domain we considered,
$x\in \left[ 10^{-8}, 10^4 \right]$, we find that $|F_0^{\rm adj}-F_0|\approx10^{-7}$, confirming the fitting quality.

\begin{table}
\caption{\label{tab:Boys_expansion}Fitting coefficients of Boys function, see Eq.~\eqref{eq:Boys_fitted_curve}.}
\begin{tabular}{cc}
\hline
\hline
\hskip0.2cm coefficient \hskip0.2cm &\hskip0.2cm fitted value \hskip0.2cm\\
\hline
    $s_1$ &  $6.70 \times10^{-5}$\\
    $B_0$ &  $1.00$\\
    $B_1$ &  $-3.33 \times10^{-1}$\\
    $B_2$ &  $9.94 \times10^{-2}$\\
    $B_3$ &  $-2.28 \times10^{-2}$\\
    $B_4$ &  $3.81 \times10^{-3}$\\
    $B_5$ &  $-3.99 \times10^{-4}$\\
    $B_6$ &  $2.15 \times10^{-5}$\\
    $s_2$ &  $6.01 \times10^{-5}$ \\
   \hline \hline
    \end{tabular}
 \end{table}

\bibliography{Bibliography}
\end{document}